\documentclass[aps,pra,showpacs,twocolumn,superscriptaddress]{revtex4}
\usepackage{graphicx,color}
\usepackage{amsmath,amsthm,amsfonts,amssymb,bm}
\usepackage{times}
\usepackage{epsf}
\usepackage[colorlinks={true}]{hyperref}

\usepackage[T1]{fontenc}
\usepackage[utf8]{inputenc}

\hypersetup{citecolor={blue}, filecolor={blue}, linkcolor={blue}, urlcolor={blue}}

\newcommand{\commentold}[1]{}
\DeclareMathSymbol{:}{\mathpunct}{operators}{"3A}
\begin{document}
\title{Witness for initial correlations among environments }
\author{F. T. Tabesh}
\affiliation{Department of Physics, University of Kurdistan, P.O.Box 66177-15175 , Sanandaj, Iran}
\affiliation{Turku Center for Quantum Physics, Department of Physics and Astronomy, University of Turku, FIN-20014 Turku, Finland}
\author{ S. Salimi}
\affiliation{Department of Physics, University of Kurdistan, P.O.Box 66177-15175 , Sanandaj, Iran}
\author{ A. S. Khorashad}
\email{a.sorouri@uok.ac.ir}
\affiliation{Department of Physics, University of Kurdistan, P.O.Box 66177-15175 , Sanandaj, Iran}
\date{\today}
\begin{abstract}
A quantum system inevitably interacts with its surroundings. In general, one does not have detailed information on an environment. Identifying the environmental features can help us to control the environment and its effects on the dynamics of an open system. Here, we consider a tripartite system and introduce a witness for the initial correlations among environments by means of the concept of the trace distance. Due to the existence of the initial environmental correlations, a tight upper bound is obtained for the growth of the trace distance of an open quantum system states. Therefore, the initial correlations among the environments subject to particular conditions can be detected by measurements on the open system.
\end{abstract}
\maketitle
\section{Introduction}
In real world, quantum systems are open systems interacting with their environments. Dynamics of an open system can be described by either Markovian or non-Markovian approach. Markovian dynamics is based on the assumptions that the coupling between the system under study and its environment is weak and that the initial system-environment (S-E) state is factorized neglecting all memory effects. Violation of any one of these conditions may lead to non-Markovian dynamics which guarantees the existence of memory effects in time evolution of an open system \cite{Breuer,Wolf}.\\
\indent As mentioned in the above, initial correlation between a system and its environment is one of the important factors to determine the Markovianity or non-Markovianity of the dynamics. Thus it plays a very important role in time evolution of an open system. If there is not any  initial correlation, the dynamics of an open system is described by a completely positive map \cite{Breuer1,Nielsen}.
In recent years, many attempts have been made to study open quantum systems with initial S-E correlations. In the presence of initial correlation,
it is shown that dynamics of an open quantum system may not be completely positive \cite{Pechukas}. In fact, it has been indicated that entangled initial states can lead to non-completely positive maps \cite{Jordan,Carteret}. In the case that quantum discord of initial states vanishes, the dynamics is described by a completely positive map \cite{Rosario}. Shabani and Lidar showed that the above-mentioned condition is not only sufficient but also necessary for complete positivity of the corresponding map\cite{Shabani1,Shabani2}. Recently, some examples were provided to show that the relation between complete positivity
and quantum discord is not generalized to all cases  \cite{Brodutch,Buscemi,Dominy}.\\
\indent The initial S-E correlations may lead to increase the trace distance over its initial value \cite{Laine}.
According to the definition of the trace distance between two arbitrary states \cite{Nielsen,Breuer}, it can be regarded as a measure for the degree of distinguishability of the two states. If the value of the trace distance during a system evolution is not constant, one can conclude that there is a flow of information between the system and its environment \cite{Laine}. A tight upper bound for its increasing has been derived which can be considered as a witness for initial S-E correlations \cite{Laine,Wismann,Smirne,Breuer2}.\\
\indent Therefore, a lot of effort has been put in to investigate the influence of initial S-E correlations on an open system dynamics. Unfortunately, a clear general  relation has not yet been found between them, and the following questions need to be answered: How do initial environmental correlations affect the dynamics of an open system? How can we obtain information about initial states of an environment?\\
\indent In this paper, we study the role of initial correlations among environments on the dynamics of an open system. For this purpose, we consider a tripartite system. In a tripartite system one can face to three scenarios: a system and two environments; two  systems and one environment; and one system, one environment and one ancilla. Here, we find an upper bound for the time evolution of the trace distance in the first scenario. When the trace distance grows above its initial value, the upper bound can be regarded as a witness for initial environmental correlations. Also, we regard some examples to illustrate the tightness of the upper bound. It should be noted that realizing initial environmental correlations may help us to characterize the environment and control its effects. In the following, we will discuss the above-mentioned questions in detail with the help of a three-qubit Heisenberg XX spin chain,  two Jaynes-Cummings
systems, two amplitude damping channels, and an experimental example. We will see that the initial correlations alter the information flow. Accordingly, initial correlations can be witnessed from the dynamical features of the open system.\\
\indent The paper is organized as follows. In Sec. II a review  of the concept of the trace distance is provided and its important role in determining the direction of information flow and also the amount of total correlations is explained. Upper bound for the growth of the distinguishability is derived  in Sec. III. In order to witness initial correlations, backflow of information is investigated for some examples in Sec. IV. The paper concludes in Sec. V.
\section{Trace distance}
The trace distance of two quantum states $\rho$  and $\sigma$ is defined as
\begin{equation}
D(\rho,\sigma) = \dfrac{1}{2}\Vert\rho-\sigma\Vert_{1},
\end{equation}
where the trace norm of an operator A is introduced by $\Vert A\Vert_{1}=Tr\vert A\vert =Tr\sqrt{A^{\dagger}A}$ \cite{Nielsen}.
It represents a metric on space of physical states, because $D\in[0,1]$, ($D(\rho,\sigma) = 0$ if and only if $\rho =\sigma$, and $D(\rho,\sigma) = 1$ if and only if $\rho$ and $\sigma$ have orthogonal supports) and  it satisfies the triangular inequality, $D(\rho,\sigma)\leq D(\rho,\tau)+ D(\tau,\sigma)$.\\
\indent The other properties of the trace distance are its subadditivity with respect to the tensor product
\begin{equation}\label{5}
D(\rho_{1}\otimes\sigma_{1},\rho_{2}\otimes\sigma_{2})\leq D(\rho_{1},\rho_{2})+D(\sigma_{1},\sigma_{2}),
\end{equation}
and its contractivity under all trace-preserving positive maps, i.e. $D(\Lambda\rho,\Lambda\sigma)\leq D(\rho,\sigma)$, where the equality  holds if $\Lambda$ is a unitary transformation. It is well known that the trace distance can be interpreted as a measure for the distinguishability of the states, therefore a trace-preserving positive map can never increase the distinguishability of any two quantum states \cite{Breuer}.\\
\indent The variation of distinguishability of two states can be considered as a witness for the flow of information in an open quantum system. Let S be an open quantum system interacting to an environment E. If $\rho^{S}_{1,2}(0)$ are two different initial states of S, their time evolutions obey  $\rho^{S}_{1,2}(t)=\Phi_{t}\rho^{S}_{1,2}(0)$, where $\Phi_{t}$ denotes the corresponding quantum dynamical map. The time variation of the trace distance is interpreted as \emph{information flow}, and is shown  by
\begin{equation}
\sigma(t) = \frac{d}{dt}D\left(\rho_{1}^{S}(t),\rho_{2}^{S}(t)\right).
\end{equation}
Positive values of $\sigma(t)$ in some time intervals correspond to information backflow from the environment to the system and the negative values indicate the information flow from the system to the environment. The quantity
\begin{equation}
I(\rho^{S}) = D\left(\rho_{1}^{S}(t),\rho_{2}^{S}(t)\right)-D\left( \rho_{1}^{S}(0),\rho_{2}^{S}(0)\right),
\end{equation}
can be regarded as a quantifier for the information exchange between an open system and its environment \cite{schmidt}.
In Eq. (4), $D\left(\rho_{1}^{S}(t),\rho_{2}^{S}(t)\right)$ can be interpreted as the information inside the system at time $t$, therefore $I(\rho^{S})$ shows the difference between the information inside the system at $t=0$ and $t$ \cite{Breuer2}. When both $I(\rho_{S})$ and $\sigma(t)$ are positive, one can obtain more information than that of the initial state of the system.\\
\indent For any state $\rho^{AB}$, the quantity $D(\rho^{AB},\rho^{A}\otimes\rho^{B})$ describes how well $\rho^{AB}$ can be distinguished from the
product state, fully uncorrelated, $\rho^{A}\otimes\rho^{B}$. Thus, $D(\rho^{AB},\rho^{A}\otimes\rho^{B})$ can be interpreted as a measure for the total amount of correlations in the state $\rho^{AB}$ \cite{Laine}. It should be mentioned that one can not recognize the correlations types by using the trace distance.\\
\indent Suppose an open system $S$ coupled to its environment $E$, with initial states $\rho_{1,2}^{SE}(0)$. Using the subadditivity and the triangular inequality of the trace distance, one can obtain the following inequality \cite{Laine}
\begin{eqnarray}
&&I(\rho^{S})=D(\rho_{1}^{S}(t),\rho_{2}^{S}(t))-D(\rho_{1}^{S}(0),\rho_{2}^{S}(0))\leq\nonumber\\
&&D(\rho_{1}^{E}(0),\rho_{2}^{E}(0))
+ \sum_{i=1}^{2}D(\rho_{i}^{SE}(0),\rho_{i}^{S}(0)\otimes\rho_{i}^{E}(0)).\nonumber\\
\end{eqnarray}
The above inequality shows an upper bound of information backflow from the environment to the system. The upper bound implies that the probable increase of the distinguishability over the initial value is due to the initial correlations in
the total initial states $\rho_{i}^{SE}(0)$ or (and) different initial states of the environment $E$ . Note that these
terms quantify both quantum and classical correlations of the total system states.\\
\indent In the next section with using the properties of the trace distance, we obtain the upper bound of the backflow of information in tripartite systems.
\begin{figure}[b]
\includegraphics[scale=0.52]{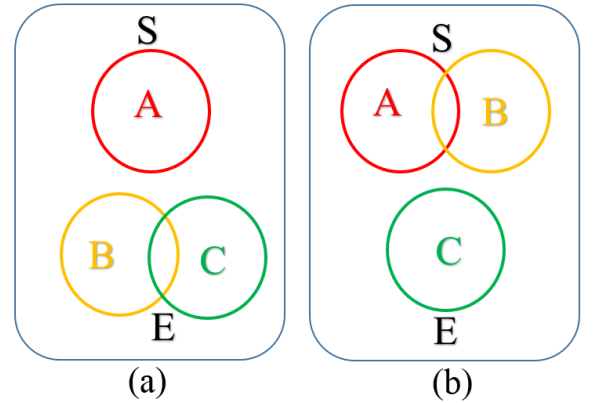}
\caption{(Color online) Schematic diagrams of a tripartite quantum system: (a) first scenario, (b) second scenario.}
\label{fig12}
\end{figure}
 \section{Dynamics of the trace distance in tripartite quantum systems}
Assume a tripartite quantum system  consists of three subsystems $A,B$ and C which can be coupled to each other.
They form an isolated system described by the initial state $\rho^{ABC}(0)$. The state of the total system at time $t$ can
be written as $\rho^{ABC}(t)=U_{t}\rho^{ABC}(0)U_{t}^{\dagger}$, where $U_{t}=\exp(\frac{-i Ht}{\hbar})$ represents the unitary time
evolution operator of the composite system with total Hamiltonian $H$. In a tripartite system one can face to three scenarios: a system
and two environments; two systems and one environment; and one system, one environment and one ancilla. The first and the second scenarios are shown in Fig. 1. Here, we investigate the first scenario.\\

  Consider the subsystem A as an open system S and the subsystems B and C as its environments. Indeed the environment E includes two subsystems B and C [see Fig. 1(a)]. Suppose two initial states $\rho^{ABC}_{1,2}(0)$ for total system, with corresponding reduced open system states $\rho^{A}_{1,2}(0)=Tr_{BC}\left(\rho^{ABC}_{1,2}(0)\right)$ and environment states $\rho^{BC}_{1,2}(0)=Tr_{A}\left(\rho^{ABC}_{1,2}(0)\right)$.
According to Eq. (5), the dynamics of the trace distance for the open system A can be written as
\begin{eqnarray}
&&D\left(\rho_{1}^{A}(t),\rho_{2}^{A}(t)\right)-D\left(\rho_{1}^{A}(0),\rho_{2}^{A}(0)\right)\leq \nonumber \\
&&\hspace{8mm}\sum_{i=1}^{2}D\left(\rho_{i}^{ABC}(0),\rho_{i}^{A}(0)\otimes\rho_{i}^{BC}(0)\right)\nonumber \\
&&\hspace{8mm}+\hspace{1mm}D\left(\rho_{1}^{BC}(0),\rho_{2}^{BC}(0)\right).
\end{eqnarray}
As stated in the introduction, our main aim is to find a witness for the initial environmental correlations, therefore, we consider the second term in the right-hand side of the above equation.
 Applying the subadditivity of the trace distance and the triangular inequality (twice) for
 $ D\left(\rho_{1}^{BC}(0),\rho_{2}^{BC}(0)\right)$, one can obtain
\begin{eqnarray}
&&D\left(\rho_{1}^{BC}(0),\rho_{2}^{BC}(0)\right)\leq \sum_{i=1}^{2}D\left(\rho_{i}^{BC}(0),\rho_{i}^{B}(0)\otimes\rho_{i}^{C}(0)\right) \nonumber \\
&&\hspace{8mm}+\hspace{1mm}D\left(\rho_{1}^{B}(0),\rho_{2}^{B}(0)\right)+D\left(\rho_{1}^{C}(0),\rho_{2}^{C}(0)\right).
\end{eqnarray}
Substituting the above inequality into Eq. (6), we find
\begin{eqnarray}
&&D\left(\rho_{1}^{A}(t),\rho_{2}^{A}(t)\right)-D\left(\rho_{1}^{A}(0),\rho_{2}^{A}(0)\right)\leq \nonumber \\
&&\hspace{5mm}\sum_{i=1}^{2}D\left(\rho_{i}^{ABC}(0),\rho_{i}^{A}(0)\otimes\rho_{i}^{BC}(0)\right) \nonumber \\
&&\hspace{3mm}+\hspace{1mm}\sum_{i=1}^{2}D\left(\rho_{i}^{BC}(0),\rho_{i}^{B}(0)\otimes\rho_{i}^{C}(0)\right) \nonumber \\
&&\hspace{3mm}+\hspace{1mm}D\left(\rho_{1}^{B}(0),\rho_{2}^{B}(0)\right)+D\left(\rho_{1}^{C}(0),\rho_{2}^{C}(0)\right),
\end{eqnarray}
where the above inequality generalizes the result of Eq. (5). This inequality shows that in the most general case an increase
of the distinguishability above its initial value implies that there must be
initial S-E correlations or initial correlations among environments or environments have  different initial states.

For the special case that there are no initial S-E correlations, the first summation in Eq. (8) vanishes and we have
\begin{eqnarray}
&&D\left(\rho_{1}^{A}(t),\rho_{2}^{A}(t)\right)-D\left(\rho_{1}^{A}(0),\rho_{2}^{A}(0)\right)\leq \nonumber \\
&&\sum_{i=1}^{2}D\left(\rho_{i}^{BC}(0),\rho_{i}^{B}(0)\otimes\rho_{i}^{C}(0)\right) \nonumber \\
&&+\hspace{1mm}D\left(\rho_{1}^{B}(0),\rho_{2}^{B}(0)\right)+D\left(\rho_{1}^{C}(0),\rho_{2}^{C}(0)\right).
\end{eqnarray}
Let us consider a further important special case, which discloses most clearly the role of initial environmental correlations,
and is obtained if we assume $\rho_{2}^{BC}(0)=\rho_{1}^{B}(0)\otimes\rho_{1}^{C}(0)$.
Therefore, the inequality in Eq. (9) is simplified to
\begin{eqnarray}
&&D\left(\rho_{1}^{A}(t),\rho_{2}^{A}(t)\right)-D\left(\rho_{1}^{A}(0),\rho_{2}^{A}(0)\right)\leq \nonumber \\
&&D\left(\rho_{1}^{BC}(0),\rho_{1}^{B}(0)\otimes\rho_{1}^{C}(0)\right),
\end{eqnarray}
where the quantity on the right-hand side of Eq. (10) can be larger than zero because of the presence of initial environmental correlations in $\rho_{1}^{BC}(0)$.
This inequality shows that any increase of the trace distance over its initial value
is a \emph{witness} for the presence of initial environmental correlations. When the inequality in Eq. (10)  becomes an equality at a certain time
$t$, we can detect the initial environmental correlations. Otherwise, the initial correlations are not transformed completely to the open system during the dynamics.\\
\indent In this step, one can ask some questions like:  where
is the rest of information stored? Has it been transformed into other
forms, or is it still frozen in bipartite environmental correlations? To answer these questions, let us recall the definition of $I_{int}(t)$ ($I_{ext}(t)$) as the information inside (outside of) the open system. Mathematically, they are written as \cite{Breuer2}
\begin{eqnarray}
&& I_{int}(t)=D(\rho_{1}^{S}(t),\rho_{2}^{S}(t)),\nonumber \\
&& I_{ext}(t) = D(\rho_{1}^{SE}(t),\rho_{2}^{SE}(t))-D(\rho_{1}^{S}(t),\rho_{2}^{S}(t)). \nonumber  \\
\end{eqnarray}
Due to the unitary dynamics of the total system, one has
\begin{eqnarray}
&& I_{ext}(0)+I_{int}(0) =I_{ext}(t)+I_{int}(t),\nonumber \\
&&I(\rho^{S})=-[I_{ext}(t)-I_{ext}(0)],\nonumber \\
\end{eqnarray}
It can clearly be seen that if $I_{int}(t)$ increases, $I_{ext}(t)$  decreases and vice versa. The second equation of Eq. (12) can be regarded as an introduction of the exchange information between the open system and the environment. Rewriting the first equation of Eq. (12) as $I_{ext}(0)=I_{ext}(t)+I_{in}(t)-I_{in}(0)$, leads us to this fact that the initially inaccessible information can  either flow to the open system or remain as external information at time $t$. With the help of Eqs. (7) and (11), one can obtain the following inequality for all $t \geq 0$:
\begin{eqnarray}
&&I_{ext}(t)\leq \sum_{i=1}^{2}D(\rho^{ABC}_{i}(t),\rho^{A}_{i}(t)\otimes\rho^{BC}_{i}(t))  \nonumber \\
&&\hspace{14mm}+ \sum_{i=1}^{2}D(\rho^{BC}_{i}(t),\rho^{B}_{i}(t)\otimes\rho^{C}_{i}(t))\nonumber \\
&&\hspace{14mm}+ D(\rho^{B}_{1}(t),\rho^{B}_{2}(t))+ D(\rho^{C}_{1}(t),\rho^{C}_{2}(t)).\nonumber \\
\end{eqnarray}
The right-hand side of the above inequality consists of six terms: The first summation  measures the total correlations between the system and the environments and the second summation measures the environmental correlations. The third and fourth terms are the trace distances of the corresponding environmental states. Thus, when $I_{ext}(t)$ grows over the initial value, $I_{ext}(0)$, the system-environment or the environment-environment correlations are created; or the environmental states become more different, implying an increase of the distinguishability of the environmental states. This demonstrates that the corresponding decrease in $I_{int}(t)$  has always an impact on degrees of freedom which are inaccessible by measurements on the open system. Conversely, if $I_{int}(t)$ starts to increase at time $t$, the corresponding decrease in $I_{ext}(t)$ implies that all kinds of correlations already exist or (and) the environmental states are different at time $t$.

 Therefore, according to Eqs. (12) and (13), the rest of the initially inaccessible information is stored in the system-environment or the environment-environment correlations,  or inside each environment.
Hence, initial environmental correlations may be transformed into other forms of bipartite or tripartite correlations.

 Here, we discuss some examples to illustrate that the inequality in Eq. (10) is tight. Suppose four qubits such that the first and second qubit are regarded as an
open system S (control qubits), and the third and fourth qubit are regarded as an
environment (target qubits), where the first (second) qubit interacts locally with the third (fourth) qubit. We first apply a controlled-NOT
gate and then a swap operation on the two qubits. Thus, the interaction is given by  unitary operator
$U=U_{1}\otimes U_{2}$, where $U_{i}=U_{swap}U_{c}$, ($i=1,2$). We consider two total initial states as
\begin{eqnarray}
&&\rho_{1}^{SE}(0)=|\varphi\rangle_{S}\langle\varphi|\otimes|\psi\rangle_{E}\langle\psi|, \nonumber\\
&&\rho_{2}^{SE}(0)=|\varphi\rangle_{S}\langle\varphi|\otimes\rho_{1}^{E_{1}}(0)\otimes\rho_{1}^{E_{2}}(0),
\end{eqnarray}
in which $|\varphi\rangle_{S}= a|00\rangle+b|11\rangle$,  $|\psi\rangle_{E}= \alpha|00\rangle+\beta|11\rangle$, where $\rho_{1}^{E}=|\psi\rangle_{E}\langle\psi|$ with $\alpha,\beta\neq 0$ and $a,b\neq 0$, and $\rho_{1}^{E_{1,2}}=Tr_{E_{2,1}}(\rho_{1}^{E})$.
The state  $\rho_{1}^{E}$ is a pure entangled state and $\rho_{2}^{E}=\rho_{1}^{E_{1}}(0)\otimes\rho_{1}^{E_{2}}(0)$ is the product of marginal states of $\rho_{1}^{E}$. For these  total states, the system states are the same.

Under the action of the unitary operator $U$ the left-hand side of Eq. (10) is found to be
\begin{eqnarray}
D(Tr_{E}(U\rho_{1}^{SE}(0)U^{\dagger}),Tr_{E}(U\rho_{2}^{SE}(0)U^{\dagger})) =|\alpha\beta|^{2}+|\alpha\beta|,\nonumber\\
\end{eqnarray}
which shows that  the trace distance of the open system states increases over its initial value. This means that the initial state of $\rho_{1}^{E}$ must  be correlated.
We also have $D(\rho_{1}^{E}(0),\rho_{1}^{E_{1}}(0)\otimes\rho_{1}^{E_{2}}(0)) =|\alpha\beta|^{2}+|\alpha\beta|$  which shows that the upper bound of the inequality in Eq. (10) is reached. Thus, the initial information in the environment state
is transferred completely to the open system by applying the the unitary operator $U$.
Now, we study a situation in which the initial environmental state has only classical correlations. Assume two total initial states as
\begin{eqnarray}
&&\rho_{1}^{SE}(0)=|\phi\rangle_{S}\langle\phi|\otimes(|\alpha|^{2}|00\rangle\langle00|+|\beta|^{2}|11\rangle\langle11|)_{E}, \nonumber\\
&&\rho_{2}^{SE}(0)=|\phi\rangle_{S}\langle\phi|\otimes\rho_{1}^{E_{1}}(0)\otimes\rho_{1}^{E_{2}}(0),
\end{eqnarray}
where $\rho_{1}^{E}$ is a purely classical state and $|\phi\rangle_{S}= a |01\rangle+ b |10\rangle$. Then one obtains
\begin{eqnarray}
&&D(Tr_{E}(U\rho_{1}^{SE}(0)U^{\dagger}),Tr_{E}(U\rho_{2}^{SE}(0)U^{\dagger})) =2|\alpha\beta|^{2},\nonumber\\
\end{eqnarray}
and the trace distance of the initial environmental states is found to be $D(\rho_{1}^{E}(0),\rho_{1}^{E_{1}}(0)\otimes\rho_{1}^{E_{2}}(0))=2|\alpha\beta|^{2}$.
We, then, see that the equality sign in Eq. (10) holds; the tightness of the bound is illustrated again.
Also, this means that the trace distance can increase even when the initial states of the environment are mixed states.\\
\indent In order to construct  initial conditions for Eq. (10), we need a second reference
state $\rho_{2}^{ABC}(0)$ whose evolution  is compared with that of
the state $\rho_{1}^{ABC}(0)$. Therefore, we regard three operators. The first one is the operator  $\textsc{\textbf{P}}$ which removes the correlations between the open system and the environments, i.e., $\textsc{\textbf{P}}(\rho_{1}^{ABC}(0))=\rho_{1}^{A}(0)\otimes\rho_{1}^{BC}(0)$.
The second one is a local trace-preserving quantum operator  generating a new state for the open system, i.e., $(\Lambda^{A}\otimes\textsc{\textbf{I}}^{BC})\circ\textsc{\textbf{P}}(\rho_{1}^{ABC}(0)) =\rho^{A}_{2}(0)\otimes\rho_{1}^{BC}(0)$.
Finally, the third one is  an operator which destroys the correlations among the environments as
\begin{eqnarray}
&&\rho^{ABC}_{2}(0)=(\textsc{\textbf{I}}^{A}\otimes\Omega^{BC})\circ(\Lambda^{A}\otimes\textsc{\textbf{I}}^{BC})\circ\textsc{\textbf{P}}(\rho^{ABC}_{1}(0))\nonumber\\
&& \hspace{14mm}=\rho^{A}_{2}(0)\otimes\rho^{B}_{1}(0)\otimes\rho^{C}_{1}(0).
\end{eqnarray}
Consequently, we have $\rho_{2}^{BC}(0)=\rho_{1}^{B}(0)\otimes\rho_{1}^{C}(0)$.\\

 In the next section, the trace distance dynamics  will be illustrated by means of a three-qubit Heisenberg XX spin chain, two Jaynes-Cummings systems, two amplitude damping channels  and an experimental example. We will see that the bound in Eq. (10) is reached for two Jaynes-Cummings systems and
the growth of the distinguishability witnesses the correlations in the initial state of the environments for these cases.
\section{ Examples}\label{III}
\subsection{ Three-Qubit  Heisenberg XX Spin Chain}
Here, interactions between three qubits  are investigated, which form  a three-qubit  Heisenberg XX spin chain \cite{Wang}.
The Hamiltonian describing the chain subject to a uniform magnetic field is
\begin{equation}\label{17}
H=\frac{J}{2}\sum_{n=1}^{3}(\sigma^{x}_{n}\sigma^{x}_{n+1}+\sigma^{y}_{n}\sigma^{y}_{n+1})+ B\sum_{n=1}^{3} \sigma^{z}_{n},
\end{equation}
where $J$ is the exchange interaction constant,
$\sigma^{\alpha}_{n}$  is the Pauli matrix corresponding to each $\alpha$ $(\alpha=x,y,z)$,  and  $B$ is the  magnitude of a uniform magnetic field. Introducing the spin raising and lowering operators of the $n${\emph{th}} qubit,  $\sigma_{n}^{\pm}=1/2 (\sigma_{n}^{x}\pm i\sigma_{n}^{y})$, the Hamiltonian can be rewritten as
\begin{equation}\label{17}
H=J\sum_{n=1}^{3}(\sigma^{+}_{n}\sigma^{-}_{n+1}+\sigma^{-}_{n}\sigma^{+}_{n+1})+ B\sum_{n=1}^{3} \sigma^{z}_{n}.
\end{equation}
Applying the periodic boundary conditions, $\sigma^{x}_{1}=\sigma^{x}_{4}$ and $\sigma^{y}_{1}=\sigma^{y}_{4}$, leads to the following
eigenvalues and eigenstates of the Hamiltonian,
\begin{eqnarray}
&&E_{0}=-E_{7}=-3B, \nonumber \\
&&E_{1}=E_{2}=-J-B, \nonumber \\
&&E_{4}=E_{5}=-J+B, \nonumber \\
&&E_{3}=2J-B, \nonumber \\
&&E_{6}=2J+B,
\end{eqnarray}
and
\begin{eqnarray}
|\psi_{0}\rangle&=&|000\rangle,\nonumber\\
|\psi_{1}\rangle&=&\frac{1}{\sqrt{3}}(e^{\frac{2 i\pi}{3}}|001\rangle+e^{\frac{-2 i\pi}{3}}|010\rangle+|100\rangle),\nonumber\\
|\psi_{2}\rangle&=&\frac{1}{\sqrt{3}}(e^{\frac{-2 i\pi}{3}}|001\rangle+e^{\frac{2 i\pi}{3}}|010\rangle+|100\rangle),\nonumber\\
|\psi_{3}\rangle&=&\frac{1}{\sqrt{3}}(|001\rangle+|010\rangle+|100\rangle),\nonumber\\
|\psi_{4}\rangle&=&\frac{1}{\sqrt{3}}(e^{\frac{2 i\pi}{3}}|110\rangle+e^{\frac{-2 i\pi}{3}}|101\rangle+|011\rangle),\nonumber\\
|\psi_{5}\rangle&=&\frac{1}{\sqrt{3}}(e^{\frac{-2 i\pi}{3}}|110\rangle+e^{\frac{2 i\pi}{3}}|101\rangle+|011\rangle),\nonumber\\
|\psi_{6}\rangle&=&\frac{1}{\sqrt{3}}(|110\rangle+|101\rangle+|011\rangle),\nonumber \\
|\psi_{7}\rangle&=&|111\rangle,
\end{eqnarray}
respectively.\\
\begin{figure}[b]
\includegraphics[scale=0.43]{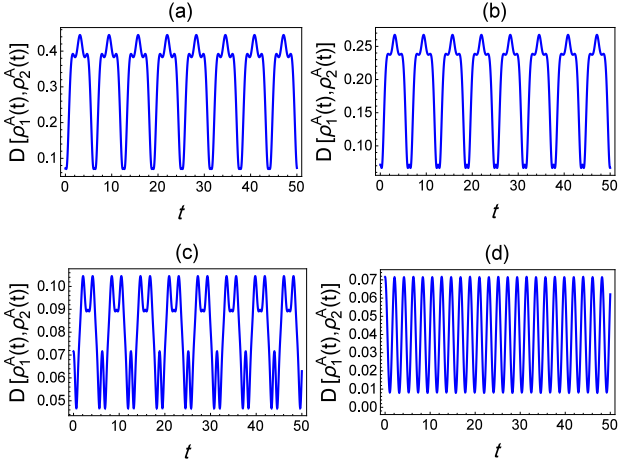}
\caption{(Color online) Plot of the trace distance of the open system A, $D(\rho^{A}_{1}(t),\rho^{A}_{2}(t))$, as
a function of time $t$, \textbf{in arbitrary units}, for the three-qubit Heisenberg XX spin chain example.  We have used $\alpha=1$ in (a) and $\alpha=0.6$ in (b). Similarly $\alpha=0.2$ in (c) and $\alpha=0$ in (d). Parameters: $f =g=1/\sqrt{2}, l=\sqrt{3/7}$, and $m=\sqrt{4/7}$. }
\label{fig12}
\end{figure}
If the normalized initial state is chosen as
\begin{equation}\label{17}
|\Psi(0)\rangle=\alpha|001\rangle+\beta|010\rangle+\gamma|100\rangle,
\end{equation}
with the help of Eqs. (21) and (22), its time evolution will be
\begin{equation}\label{17}
|\Psi(t)\rangle=a(t)|001\rangle+b(t)|010\rangle+c(t)|100\rangle,
\end{equation}
where
\begin{equation}\label{17}
\begin{split}
a(t)= \frac{1}{3}( e^{it(J+B)}(2\alpha-\beta-\gamma)+ K(t)),\\
b(t)=\frac{1}{3}( e^{it(J+B)}(2\beta-\alpha-\gamma)+ K(t)),\\
c(t)=\frac{1}{3}( e^{it(J+B)}(2\gamma-\alpha-\beta)+ K(t)),
\end{split}
\end{equation}
in which $K(t)=e^{-it(2J-B)} (\alpha+\beta+\gamma)$.\\
\indent As a different case, one can assume that there are two excitations in the total system. Thus, the initial state is defined as
\begin{equation}\label{17}
|\Phi(0)\rangle=\alpha_{1}|110\rangle+\beta_{1}|101\rangle+\gamma_{1}|011\rangle,
\end{equation}
and its time evolution is determined by
\begin{equation}\label{17}
|\Phi(t)\rangle=a_{1}(t)|110\rangle+b_{1}(t)|101\rangle+c_{1}(t)|011\rangle,
\end{equation}
where
\begin{equation}\label{17}
\begin{split}
a_{1}(t)= \frac{1}{3}(e^{-it(-J+B)}(2\alpha_{1}-\beta_{1}-\gamma_{1})+Z(t)),\\
b_{1}(t)=\frac{1}{3}(e^{-it(-J+B)}(2\beta_{1}-\alpha_{1}-\gamma_{1})+ Z(t)),\\
c_{1}(t)=\frac{1}{3}(e^{-it(-J+B)}(2\gamma_{1}-\alpha_{1}-\beta_{1})+ Z(t)),
\end{split}
\end{equation}
in which $Z(t)=e^{-it(2J+B)}(\alpha_{1}+\beta_{1}+\gamma_{1})$.\\
\indent In order to show the influence of the initial environmental correlations on the trace distance dynamics, we illustrate three situations. Note that we regard the first qubit as an open system S and the other two qubits as its environment E [see Fig. 1(a)].\\
\indent i) For the first case, let us assume two environmental states such that only one of them has initial correlations. Hence, we regard the total initial states as
\begin{equation}\label{17}
\rho_{1}(0)=|\varphi\rangle_{A}\langle\varphi|\otimes\left(\frac{1-\alpha}{4}I+\alpha|\psi^{-}\rangle\langle\psi^{-}|\right)_{BC},
\end{equation}and
\begin{equation}\label{17}
\rho_{2}(0)=|\phi\rangle_{A}\langle\phi|\otimes\frac{1}{2}I_{B}\otimes\frac{1}{2}I_{C},
\end{equation}
where $\rho^{BC}_{1}(0)$ is a Werner state, $|\varphi\rangle_{A}=f|0\rangle+g|1\rangle$, $|\phi\rangle_{A}=l|0\rangle+m|1\rangle$, and $|\psi^{-}\rangle=\frac{1}{\sqrt{2}}(|01\rangle-|10\rangle)$.

\indent For these states, we have  $D\left(\rho^{B}_{1}(0),\rho^{B}_{2}(0)\right)=0$, $D\left(\rho^{C}_{1}(0),\rho^{C}_{2}(0)\right)=0$, $D\left(\rho^{BC}_{2}(0),\rho^{B}_{2}(0)\otimes\rho^{C}_{2}(0)\right)=0$, and  initial S-E correlations are zero. According to Eq. (10), the upper bound of the increase of the trace distance is restricted to the initial correlations among the environments in $\rho_{1}(0)$.

\indent In order to calculate the trace distance dynamics of the open system A, we find the time evolution of these total states from Eqs. (24) and (27). Then, with tracing over the environments (B+C), the reduced open system dynamics can be obtained. The behavior of the trace distance of $\rho^{A}$ as
a function of  $t$ is plotted in Fig. 2. Different initial states are considered with parameters $f =g=1/\sqrt{2}, l=\sqrt{3/7}$, and $m=\sqrt{4/7}$ . In Figs. 2(a), (b), (c), and (d) the values of  $\alpha$ are assumed to be $1, 0.6, 0.2$, and $0$, respectively.\\
\indent In Fig. 2(a), initial state of the environments in $\rho_{1}(0)$ is defined by a Bell state ($\alpha=1$), a maximally entangled state. As can be seen, the trace distance begins to increase after the initial time. This means that an amount of the initial environmental correlations flows to the  open system from the beginning of the dynamics. Furthermore, it has a periodic behavior during the dynamics. In Fig. 2(b), the initial state of the environments in $\rho_{1}(0)$ is not maximally entangled state and it is characterized by $\alpha=0.6$. From the figure one can see  that the amount of information backflow is reduced by decreasing the initial environmental correlations although the dynamics behaviour is similar to Fig. 2(a).

\indent The value $\alpha=0.2$ is used in Fig. 2(c), where the amount of quantum initial correlations decreases such that the amount of entanglement is zero but the amount of discord is not. We remark that the trace distance starts decreasing already at the initial time then it begins to grow at a later time.
In Fig. 2(d), the initial state of the environments in $\rho_{1}(0)$ is given by $\alpha=0$. Note that in this case the trace distance does not increase over its initial value since there is no initial correlation between environments. \\
\indent In brief, Fig. 2 shows the effect of initial correlations among the environments on the trace distance dynamics of the open system. We conclude, for this example, that the amount of the information backflow from the environments to the open system is increased by  increasing initial quantum correlations among the environments and it can lead to increase distinguishability over its initial value.
In situations investigated in Fig. 2, the maximum amount of the trace distance as a function of time is
not equal to the upper bound given by Eq. (10). This means that the information initially inaccessible to the open system
has not been transferred completely to it during the dynamics.

\indent ii) For the second situation, let us study an example in which the both initial environmental states have quantum correlations. In this and the next example, we  use Eq. (9) to witness the initial environmental correlations. The total initial states can be taken as
\begin{eqnarray}
&& \rho_{1}(0)=|\varphi\rangle_{A}\langle\varphi|\otimes(\frac{1-\alpha_{1}}{4}I+\alpha_{1}|\psi^{-}\rangle\langle\psi^{-}|)_{BC},\nonumber \\
&& \rho_{2}(0)=|\phi\rangle_{A}\langle\phi|\otimes(\frac{1-\alpha_{2}}{4}I+\alpha_{2}|\psi^{-}\rangle\langle\psi^{-}|)_{BC}. \nonumber  \\
\end{eqnarray}
In Fig. 3(a) the dynamics of $D(\rho_{1}^{A}(t),\rho_{2}^{A}(t))$ is shown for $\alpha_{1}=1$ and $\alpha_{2}=0.6$. If this figure is compared with Figs. 2(a) and (b), one realizes that the both quantum correlations  have destructive effect on the distinguishability of the open system states which means that the amount of information flowing to the system is little. Equation
\begin{eqnarray}
D(\rho^{BC}_{1}(0),\rho^{BC}_{2}(0))=\frac{3}{4}|\alpha_{1}-\alpha_{2}|,\nonumber \\
\end{eqnarray}
implies that the maximum information outside of the open system can be obtained for $\alpha_{1}=1$, $\alpha_{2}=0$, and $\alpha_{1}=0$, $\alpha_{2}=1$. Therefore, the more difference among the initial quantum correlations (initial environmental states), the more information is initially stored outside of the open system and as a result the distinguishability of the open system states increases over its initial value. Actually, in order to have more information flowed to the open system, the difference among the initial quantum correlations must be more. A maximally entangled state and a product state are suitable candidates for this purpose (for the initial environmental states).

\indent iii) For the third one, let us consider a situation in which there is quantum correlation in one of the two initial environmental states and classical correlation in the other. An example for this case can be
\begin{eqnarray}
&&\rho_{1}(0)=|\varphi\rangle_{A}\langle\varphi|\otimes(\frac{1-\alpha}{4}I+\alpha|\psi^{-}\rangle\langle\psi^{-}|)_{BC},\nonumber \\
&&\rho_{2}(0)=|\phi\rangle_{A}\langle\phi|\otimes\frac{1}{2}(|00\rangle\langle 00|+|11\rangle\langle 11|). \nonumber  \\
\end{eqnarray}
Fig. 3(b) shows the time behavior of the trace distance of the open system states for $\alpha=1$. As can be seen, the maximum value of the distingushability is 0.75. Comparing Fig. 3(b) with Fig. 3(a) and Fig. 2, leads us to this fact that maximal classical and quantum correlations are the best choice for obtaining maximum inaccessible initial information. Thus, the states with the above-mentioned properties have effective influence on the growth of the distinguishability of the open system states. This is confirmed by
\begin{eqnarray}
D(\rho^{BC}_{1}(0),\rho^{BC}_{2}(0))=\frac{1+\alpha}{2},\nonumber \\
\end{eqnarray}
showing that the information outside of the open system gets its maximum value when $\alpha=1$ (maximally entangled state).\\
\indent Studying the above examples shows that whenever  more distinguishable the environmental states are, the more information is stored outside of the open system; and returned information to the open system is maximum if  there are initial classical and quantum correlations. Although the presence of quantum correlations in the both of the initial environmental states has destructive effect on the growth of the distiguishability of the open system states, initial quantum-classical correlations constructively affect the distinguishability.\\
\indent In the next subsection we introduce two Jaynes-Cummings systems by which one can show that the inequality in Eq. (10) is tight.
\begin{figure}[t]
\includegraphics[scale=0.49]{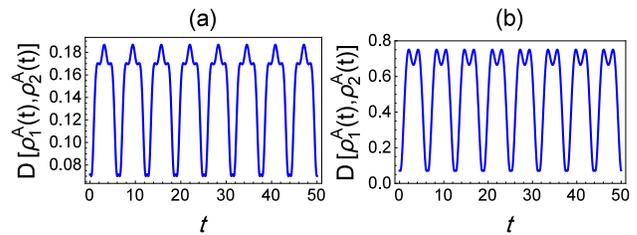}
\caption{(Color online) Plot of  $D(\rho_{1}^{A}(t),\rho_{2}^{A}(t))$, for the three-qubit Heisenberg XX spin chain example, as
a function of time $t$, \textbf{in arbitrary units}. We have used $\alpha_{1}=1$ and $\alpha_{2}=0.6$ in (a) and $\alpha=1$ in (b).  }
\label{fig12}
\end{figure}

\subsection{  Two Jaynes-Cummings systemes}
i) Suppose that one provides two Jaynes-Cummings systems in which each atom is locally coupled to a single-mode field. In this case, the open system of the tripartite system is assumed to include two atoms and each field is regarded as an environment. The total Hamiltonian is given by
\begin{eqnarray}\label{17}
&&H=H^{(1)}+H^{(2)},
\end{eqnarray}
where
\begin{eqnarray}
&&H^{(j)}=\omega_{0}^{j}\sigma_{+}^{j}\sigma_{-}^{j}+\omega^{j}b^{j \dag}b^{j}+g^{j}(\sigma_{+}^{j}b^{j}+\sigma_{-}^{j}b^{j \dag}),\nonumber
\end{eqnarray}
in which $\sigma_{+}^{j}(\sigma_{-}^{j})$ is the raising (lowering) operator of the $j$th atom, $b^{j\dag}$ $(b^{j})$
is the creation (annihilation) operator of the $j${th} field,  $\omega_{0}^{j}$ is the frequency of the $j${th} atom, $\omega^{j}$ is the frequency of the $j${th} field,  and $g^{j}$ is the coupling constant between the $j${th} atom and the $j${th} field ($j=1,2$).
In the interaction picture the Hamiltonian takes the following form
\begin{equation}\label{17}
H_{I}^{(j)}=g^{j}(\sigma_{+}^{j}b^{j} e^{i \Delta^{j}(t)}+\sigma_{-}^{j}b^{j \dag}e^{-i \Delta^{j}(t)}),
\end{equation}
where $\Delta^j=\omega_{0}^j-\omega^j$ is the detuning between the $j$th atom and the $j$th  field. Let us assume  that $b^{1}=b^{2}=b$, $g^{1}=g^{2}=g$, $\omega_{0}^{1}=\omega_{0}^{2}=\omega_{0}$, and $\omega^{1}=\omega^{2}=\omega$,  hence, $\Delta=\Delta^1=\Delta^2= \omega_{0}-\omega$.
The  local time evolution operator in the interaction picture can be written as
\begin{equation}\label{covariance matrix}
U^{(j)}(t)= \begin{pmatrix}
         c(\hat{n}+1,t) & d(\hat{n}+1,t)b \\
        -b^{\dag}d^{\dag}(\hat{n}+1,t) & c(\hat{n},t) \\
\end{pmatrix},
\end{equation}
where
\begin{eqnarray}
&&c(\hat{n},t)=e^{i \Delta t/2} \left[\cos\left(\Omega(\hat{n})\frac{t}{2}\right)-i\frac{\Delta}{\Omega(\hat{n})}\sin\left(\Omega(\hat{n})\frac{t}{2}\right)\right],\nonumber\\
&&d(\hat{n},t)=- i e^{i \Delta t/2}  \frac{2g}{\Omega(\hat{n})}\sin\left(\Omega(\hat{n})\frac{t}{2}\right),\nonumber\\
\end{eqnarray}
in which $\Omega(\hat{n})=\sqrt{\Delta^{2}+4 g^{2} \hat{n}}$ \cite{puri}.

The $i$th reduced density matrix of the system at time $t$ can be written as
\begin{eqnarray}\label{17}
&&\rho^{S}_{i}(t)=\nonumber\\
&&Tr_{E}\left[U^{(1)}(t)\otimes U^{(2)}(t)\left(\rho_{i}(0)\right)U^{(1)\dag}(t) \otimes U^{(2)\dag}(t)\right],\nonumber\\
\end{eqnarray}
where $\rho_{i}(0)$ is the $i$th initial state of the total system and it is assumed to be a product state as $\rho_{i}(0)=\rho^{S}(0)\otimes\rho^{BC}_{i}(0)\quad (i=1,2,3)$. Let the initial state of the open system be $\rho^{S}(0)=|ee\rangle\langle ee|$.  The first environmental initial state is taken as
\begin{eqnarray}\label{17}
\rho^{BC}_{1}(0)=
(\alpha|0,n\rangle+\beta|n,0\rangle)(\alpha^{\ast}\langle 0,n|+\beta\langle n,0|),\nonumber\\
\end{eqnarray}
which shows entanglement among the environments. The second one is built by the marginal states of the first environmental initial state as
 $\rho^{BC}_{2}(0)=\rho^{B}_{1}(0)\otimes \rho^{C}_{1}(0)$ which is obtained as
\begin{eqnarray}\label{17}
&&\rho^{BC}_{2}(0)=|\alpha|^{4}|0,n\rangle\langle 0,n|+|\beta|^{4}|n,0\rangle\langle n,0|\nonumber\\
&&+|\alpha|^{2}|\beta|^{2}\left( |0,0\rangle\langle 0,0|+|n,n\rangle\langle n,n| \right).\nonumber\\
\end{eqnarray}
Finally, the third state is chosen to be a classically correlated state
\begin{eqnarray}\label{17}
&&\rho^{BC}_{3}(0)=|\alpha|^{2}|0,0\rangle\langle 0,0|+|\beta|^{2}|n,n\rangle\langle n,n|.\nonumber\\
\end{eqnarray}
\begin{figure}[t]
\includegraphics[scale=0.43]{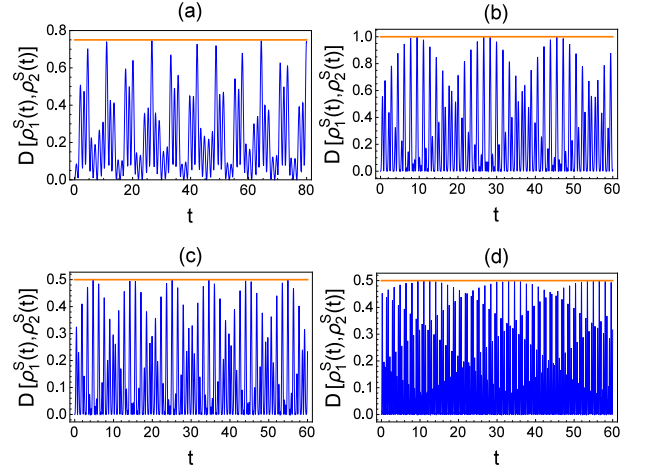}
\caption{(Color online) The trace distance dynamics of the open system for the Jaynes-Cummings example as a function of time $t$, \textbf{in arbitrary units}, and $g = 1$  (a)  $D(\rho^{S}_{1}(t),\rho^{S}_{2}(t))$, with $\Delta=0.1$ and $n=1$  (b) $D(\rho^{S}_{1}(t),\rho^{S}_{3}(t))$, with $n=7$ and  $\Delta=0$ (c) and (d) $D(\rho^{S}_{3}(t),\rho^{S}_{2}(t))$, with $\Delta=0$,  $n=10$ and $n=50$, respectivrly. The horizontal line denotes the upper bound of Eq. (10) (Eq. (5)) in figures (a), (c), and (d) ((b)).}
\label{fig12}
\end{figure}
\indent Substituting the above three initial states into Eq. (39) and taking into account Eqs. (37) and (38), one can obtain the dynamics of the open system. The trace distance dynamics of the open system states is plotted in Fig. 4 for $g=1$ and $\alpha=\beta=1/\sqrt{2}$. Fig. 4(a) shows the time behavior of $D(\rho^{S}_{1}(t),\rho^{S}_{2}(t))$ for $n=1$ and $\Delta=0.1$. The distinguishability value of the initial environmental states is $D(\rho^{BC}_{1}(0),\rho^{BC}_{2}(0))=0.75$ in which $\rho^{BC}_{1}(0)$ is maximally entangled state and $\rho^{BC}_{2}(0)$ is a product one. As can be seen, the total initial entanglement among two modes flows to the system at some points of time. It actually shows that the bound of the inequality in Eq. (10) is tight. \\
\indent For $n=7$ and  $\Delta=0$, $D(\rho^{S}_{1}(t),\rho^{S}_{3}(t))$ is plotted against time in Fig. 4(b). In this case, $D(\rho^{BC}_{1}(0),\rho^{BC}_{3}(0))=1$, and $D(\rho^{BC}_{1}(0),\rho^{BC}_{2}(0))+D(\rho^{BC}_{2}(0),\rho^{BC}_{3}(0))=1.25$ which is greater than $1$.
According to Eqs. (40) and (42), one can realize that there is quantum correlation in $\rho^{BC}_{1}(0)$, whereas, $\rho^{BC}_{3}(0)$ is a classically correlate state.
This is an example for which the inequality in Eq. (5) is tight but  the one in Eq. (9) is not. The plot shows that the trace distance reaches $1$ at some values of time, and therefore, the open system becomes completely distingushable in those values of time.\\
\indent In Figs. 4(c) and 4(d), $D(\rho^{S}_{3}(t),\rho^{S}_{2}(t))$ is  depicted for $\Delta=0$, and for two values of $n$, $10$ and $50$, respectively. The trace distance of the initial environmental states is  $D(\rho^{BC}_{3}(0),\rho^{BC}_{2}(0))=0.5$. As can be seen the upper bound is reached for both values of $n$.\\
\indent In summary, Fig. 4 shows that the upper bound is tight and the distingushability reaches $1$ when there are initial quantum-classical correlations among the fields. Furthermore, it indicates that initial quantum correlations make the trace distance increase more than classical correlations do.\\
\indent ii) Let us assume an example showing the tightness of the upper bound for classical states. To this aim, the total  initial states are taken as
\begin{eqnarray}\label{17}
&&\rho_{1}(0)=|ee\rangle\langle ee|\otimes \frac{1}{2}(|\beta,-\beta\rangle\langle \beta,-\beta|+|-\beta,\beta\rangle\langle -\beta,\beta|),\nonumber\\
&&\rho_{2}(0)=|ee\rangle\langle ee|\otimes \frac{1}{2}(|\beta\rangle\langle \beta|+|-\beta\rangle\langle -\beta|)\nonumber\\
&&\hspace{24mm}\otimes \frac{1}{2}(|-\beta\rangle\langle -\beta|+|\beta\rangle\langle\beta|),\nonumber\\
\end{eqnarray}
in which $|\beta\rangle= e^{-|\beta|^{2}/2}\sum_{n=0}^{\infty}\frac{\beta^{n}}{\sqrt{n!}}|n\rangle$ is a coherent state with mean number of photons as $\langle n\rangle=|\beta|^{2}$. It is well known that the coherent state does always have minimum uncertainty and resembles a classical state. Substituting the initial states into Eq. (39), one can obtain $D(\rho^{S}_{1}(t),\rho^{S}_{2}(t))$. For $\Delta=0$ and $g=1$, the trace distance dynamics is plotted for $|\beta|^{2}=100$ and  $|\beta|^{2}=200$, in Figs. 5(a) and 5(b), respectively. One can see that as the average number of photons increases, the initial total classical correlation among the modes is detected at a given time. Therefore, the bound is tight for classical state and our witness can be applied for those states.\\
\indent The above two examples indicate that one can detect the initial quantum and classical correlations among two fields by studying the dynamics of the trace distance of the system states.\\
In the following the witness can be applied for a dissipative dynamics. For this purpose a discussion on  amplitude damping channels is provided.
\begin{figure}[t]
\includegraphics[scale=0.52]{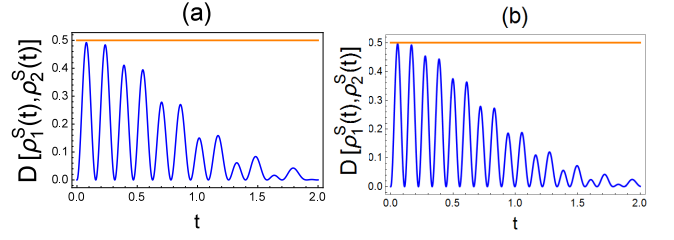}
\caption{(Color online)  Plot of $D(\rho^{S}_{1}(t),\rho^{S}_{2}(t))$ as a function of time $t$, \textbf{in arbitrary units}, and $\Delta=0$ and  $g = 1$, for the Jaynes-Cummings example.  In both figures the horizontal line marks the upper bound of Eq. (10). (a)$|\beta|^{2}=100$, for this value the bound is not tight (b) $|\beta|^{2}=200$, as can be seen total initial classical correlation can be observed in a given time.}
\label{fig12}
\end{figure}
\subsection{ Amplitude damping model}
Here, we consider an open system consisting of two atoms locally interacting with amplitude damping reservoir. The Hamiltonian $H$ of the whole system is defined as
\begin{eqnarray}\label{17}
H=H^{(1)}+H^{(2)},\nonumber\\
\end{eqnarray}
where
\begin{eqnarray}\label{17}
H^{(i)}=\omega_{0}^{i}\sigma_{+}^{i}\sigma_{-}^{i}+\sum_{k=0}\omega_{k}^{i}b_{k}^{i \dag }b_{k}^{i}+\sum_{k=0}g_{k}^{i}(\sigma_{+}^{i}b_{k}^{i}+\sigma_{-}^{i}b_{k}^{ i \dag});\nonumber\\
\nonumber
\end{eqnarray}
in which $b_{k}^{i\dag}$ ($b_{k}^{i}$) is the creation (annihilation) operator corresponding to the $k$th mode of the $i${th} reservoir, $\omega_{k}^{i}$  is the frequency of the $k$th mode of the $i${th} reservoir, $\omega_{0}^{i}$ is the frequency related to the transition energy of the $i${th} atom, $ g_{k}^{i}$ is the coupling constant between the $i${th} atom and the $k$th mode of the $i${th} reservoir, and $\sigma_{+}^{i}(\sigma_{-}^{i})$ is the raising (lowering) operator of the $i${th} atom ($i=1,2$).
We suppose that the two atoms have the same transition energy and the same coupling to the reservoirs. Furthermore, we assume that the both reservoirs have the same Lorentz spectral density\cite{ban,ban2}.\\
\indent In order to introduce an initial state, let us define the vacuum state as $|\textbf{0}\rangle=|0_{1}0_{2}...0_{k}...\rangle$, therefore a first excited state is  $|\textbf{1}_{k}\rangle=|0_{1}0_{2}...0_{k-1}1_{k}0_{k+1}...\rangle$ in which $|1_{k}\rangle=b_{k}^{\dag}|0_{k}\rangle$. It is obvious that the both states are orthogonal, i.e.  $\langle \textbf{0}|\textbf{1}\rangle=0$. Total initial state is assumed to be a superposition of two states. In one state, atoms are in a Bell state and the reservoirs are in the vacuum states. The other one is that the two qubits are in the ground states and one of the two reservoirs has only one excitation.
Thus the initial state of the total system is written as
\begin{eqnarray}\label{17}
&&|\psi(0)\rangle=c_{eg}(0)|\psi_{+}\rangle|\textbf{0},\textbf{0}\rangle+\sum_{k}c_{k}(0)|g,g\rangle\otimes|\textbf{1}_{k},\textbf{0}\rangle \nonumber\\ &&\hspace{14mm}+\sum_{k}d_{k}(0)|g,g\rangle\otimes|\textbf{0},\textbf{1}_{k}\rangle;\nonumber\\
\end{eqnarray}
where $|\psi_{+}\rangle=\frac{1}{\sqrt{2}}(|e,g\rangle+|g,e\rangle)$ is a  Bell state and $|g\rangle$ ($|e\rangle$)refers to the ground (excited) state of each atom. The normalization condition for $|\psi(0)\rangle$ is $|c_{eg}(0)|^{2}+|\sum_{k}c_{k}(0)|^{2}+|\sum_{k}d_{k}(0)|^{2}=1$. In the case $c_{k}(0)=d_{k}(0)$, the state of the whole system at time $t$ is written as
\begin{eqnarray}\label{17}
&&|\psi(t)\rangle=c_{eg}(t)|\psi_{+}\rangle|\textbf{0},\textbf{0}\rangle+\sqrt{1-|c_{eg}(t)|^{2}}|g,g\rangle\otimes|\psi_{+}^{t}\rangle, \nonumber\\ \end{eqnarray}
where
\begin{eqnarray}\label{17}
&&c_{eg}(t)=h_{1}(t)c_{eg}(0)+h_{2}(t)\sqrt{1-|c_{eg}(0)|^{2}},  \nonumber\\
\end{eqnarray}
in which
\begin{eqnarray}\label{17}
&&h_{1}(t)= e^{-\frac{1}{2}\lambda t}\left[\cosh\left(\frac{\lambda a}{2}t\right)+\frac{1}{a}\sinh\left(\frac{\lambda a}{2}t\right)\right], \nonumber\\
&&h_{2}(t)=-i e^{-\frac{1}{2}\lambda t}\left[\sqrt{\frac{1}{a^{2}}-1}\sinh\left(\frac{\lambda a}{2}t\right)\right],
\end{eqnarray}
with $a=\sqrt{1-2\frac{\gamma}{\lambda}}$, where $\gamma$ is connected to the time scale of the system and $\lambda$ is coupling spectral width.
Also in Eq. (46), $|\psi_{+}^{t}\rangle$ is a Bell state of the two reservoirs which is  $\frac{1}{\sqrt{2}}(|\textbf{1}^{t},\textbf{0}\rangle+|\textbf{0},\textbf{1}^{t}\rangle)$. The first excitation state of each reservoir depends on time as
\begin{eqnarray}\label{17}
&&|\textbf{1}^{t}\rangle= \frac{1}{\sqrt{\sum_{k}|c_{k}(t)|^{2}}}\sum_{k}c_{k}(t)|\textbf{1}_{k}\rangle ,\nonumber\\
\end{eqnarray}
which  is  normalized,  $\langle \textbf{1}^{t}| \textbf{1}^{t}\rangle=1$, and  orthogonal to $|\textbf{0}\rangle$, $\langle \textbf{0}| \textbf{1}^{t}\rangle=0$. \\
\indent  An initial state of the whole system can be obtained if one has
$c_{eg}(0)=0$, $|\sum_{k}c_{k}(0)|^{2}=1/2$, and $|\textbf{1}\rangle= (\sum_{k}|c_{k}(0)|^{2})^{-1/2}\sum_{k}c_{k}(0)|\textbf{1}_{k}\rangle$, which result in $|\psi_{1}(0)\rangle=|gg\rangle \otimes\frac{1}{\sqrt{2}}(|\textbf{1},\textbf{0}\rangle+|\textbf{0},\textbf{1}\rangle)$. Therefore, the initial environmental state is an entangled state. Regarding the above assumptions,  the state of the atoms at time $t$ is
\begin{eqnarray}\label{17}
&&\rho^{S}_{1}(t)=|h_{2}(t)|^{2}|\psi_{+}\rangle\langle\psi_{+}|+(1-|h_{2}(t)|^{2})|gg\rangle\langle gg|.  \nonumber\\
\end{eqnarray}
Another initial state is assumed to be a product state as
\begin{eqnarray}\label{17}
&&\rho_{2}(0)=|gg\rangle\langle gg|\otimes \rho^{B}_{1}(0)\otimes\rho^{C}_{1}(0),  \nonumber\\
\end{eqnarray}
in which
\begin{eqnarray}\label{17}
&&\rho^{B}_{1}(0)=\rho^{C}_{1}(0)=\frac{1}{2}(|\textbf{0}\rangle\langle \textbf{0}|+|\textbf{1}\rangle\langle \textbf{1}|).\nonumber\\
\nonumber
\end{eqnarray}
Regarding  Eq. (46) and the corresponding equations in \cite{ban}, the reduced density matrix of the atoms gets the following form
\begin{equation}\label{covariance matrix}
\rho^{S}_{2}(t)= \frac{1}{4}\begin{pmatrix}
         \rho_{ee}(t) & 0& 0&0 \\
         0 & \rho_{eg}(t) &0&0\\
         0 &0 & \rho_{ge}(t)&0\\
       0 &0 &0 &\rho_{gg}(t) \\
\end{pmatrix},
\end{equation}
with
\begin{eqnarray}\label{17}
&&\rho_{ee}(t)=|h_{2}^{2}(t)|^{2},  \nonumber\\
&&\rho_{eg}(t)=\rho_{ge}(t)=|h_{2}(t)|^{2}(2-|h_{2}(t)|^{2}), \nonumber\\
&&\rho_{gg}(t)=(2-|h_{2}(t)|^{2})^{2}.\nonumber\\
\end{eqnarray}
\begin{figure}[t]
\includegraphics[scale=0.39]{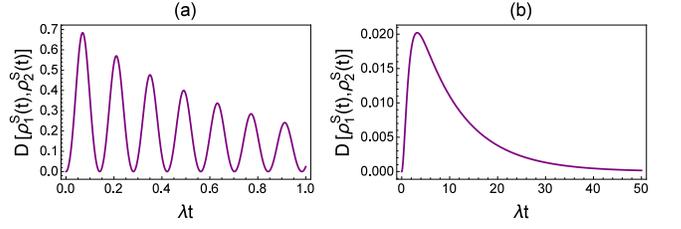}
\caption{(Color online) Plot of $D(\rho^{S}_{1}(t),\rho^{S}_{2}(t))$ as a function of scaled time $\lambda t$ for the amplitude damping example (a) local non-Markovian dynamics ($\gamma/\lambda=1000$), (b) local Markovian dynamics ($\gamma/\lambda=0.1$).}
\label{fig12}
\end{figure}
\indent For two amplitude damping channels, the trace distance dynamics of the open system (the atoms), $D(\rho^{S}_{1}(t),\rho^{S}_{2}(t))$, is plotted against time    ($\lambda t$) for  $\gamma/\lambda=1000$ (local non-Markovian dynamics) in Fig. 6(a) and  for $\gamma/\lambda=0.1$ (local Markovian  dynamics) in Fig. 6(b). Here, the value of the initial environment-environment correlation is $D(\rho^{BC}_{1}(0),\rho^{BC}_{2}(0))=0.75$, and as can be seen in Fig. 6  the upper bound is not reached for the both cases. It is clear from Fig. 6(a), that the trace distance damply oscillates as a function of time, however, no oscillation can be seen in Fig. 6(b). The oscillation of the trace distance in Fig. 6(a) shows that information repeatedly exchanges between the system and environments; and comparing the plot in Fig. 6(a) with that in Fig. 6(b) indicates that the value of the exchanged information in the first case is greater than that in the second case. It should be mentioned that in the case of initial classical correlation, our calculations show that the inequality in Eq. (10) is not tight.\\
\indent As a final example, in the next subsection, let us consider an experimental one which has been introduced by other authors\cite{Laine1}.
\subsection{ Experimental example}
As an experimental example, we consider two entangled photons whose polarization degrees of freedom locally interact with their frequency degrees of freedom. The polarization degrees of freedom of the photons are regarded as an open system and their frequency degrees of freedom form two environments\cite{Laine1}.
The Hamiltonian of the local interaction is defined by
\begin{equation}\label{17}
H_{i}=- \int d\omega_{i}\omega_{i}(n_{V}|V\rangle\langle V|+n_{H}|H\rangle\langle H|)|\omega_{i}\rangle\langle\omega_{i}|,
\end{equation}
where  $|H\rangle$ ($|V\rangle$) and  $|\omega_{i}\rangle$ indicate the state of a photon with horizontal (vertical) polarization and frequency $\omega_{i}$, respectively.
The refraction index for photon with polarization $H$ ($V$) is signified by $n_{H}$ ($n_{V}$). We assume the total initial state as
\begin{equation}\label{17}
|\Psi(0)\rangle=|\psi^{12}\rangle\otimes\int d\omega_{1}d\omega_{2} g(\omega_{1},\omega_{2})|\omega_{1},\omega_{2}\rangle,
\end{equation}
where $|\psi^{12}\rangle=a|HH\rangle+b|HV\rangle+c|VH\rangle+d|VV\rangle$ and $g(\omega_{1},\omega_{2})$ denotes the probability amplitude for the first photon to have frequency $\omega_{1}$ and the second photon to have frequency $\omega_{2}$, with the corresponding joint probability distribution
$P(\omega_{1},\omega_{2})=|g(\omega_{1},\omega_{2})|^{2}$.\\
\indent Due to the initial product system-environments state, the time evolution of the open system can be described as $\rho^{S}(t)=\Phi_{t}^{12}(\rho^{S}(0))$, $\rho^{S}(0)=|\psi^{12}\rangle\langle\psi^{12}|$, where  $\Phi_{t}^{12}$ is a dynamical map which maps the
initial polarization state to the polarization state at time $t$. The state of the open system at time $t$ is given by
\begin{equation}\label{covariance matrix}
\rho^{S}(t)= \begin{pmatrix}
          |a|^{2} & ab^{\ast}\kappa_{2}(t) & ac^{\ast}\kappa_{1}(t) & ad^{\ast}\kappa_{12}(t) \\
          b a^{\ast}\kappa^{\ast}_{2}(t) & |b|^{2} & bc^{\ast}\Lambda_{12}(t) &bd^{\ast}\kappa_{1}(t) \\
         c a^{\ast}\kappa^{\ast}_{1}(t) &bc^{\ast}\Lambda^{\ast}_{12}(t) & |c|^{2} & cd^{\ast}\kappa_{2}(t) \\
        d a^{\ast}\kappa^{\ast}_{12}(t) &db^{\ast}\kappa^{\ast}_{1}(t) &dc^{\ast}\kappa^{\ast}_{2}(t) & |d|^{2} \\
\end{pmatrix},
\end{equation}
in which $\kappa_{1}(t)=G(\Delta n t_{1},0)$, $\kappa_{2}(t)=G(0,\Delta n t_{2})$, $\kappa_{12}(t)=G(\Delta n t_{1},\Delta n t_{2})$, and $\Lambda_{12}(t)=G(\Delta n t_{1},-\Delta n t_{2})$, where
\begin{equation}\label{17}
G(\tau_{1},\tau_{2})=\int d\omega_{1}d\omega_{2} P(\omega_{1},\omega_{2})e^{-i(\omega_{1}\tau_{1}+\omega_{2}\tau_{2})}
\end{equation}
is the Fourier transform of the joint probability distribution and $\Delta n= n_{V}-n_{H}$.
\begin{figure}[t]
\includegraphics[scale=0.7]{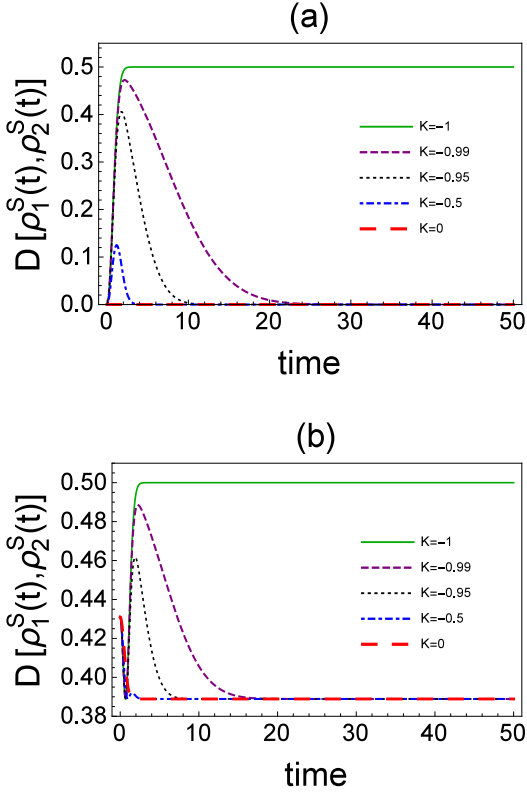}
\caption{(Color online) The trace distance dynamics of the open system in the experimental example, $D(\rho^{S}_{1}(t),\rho^{S}_{2}(t))$,  for different values of K, as a function of time, \textbf{in arbitrary units}. The time scale is $\sqrt{C_{11}}\Delta n t$.}
\label{fig12}
\end{figure}
\indent Note that the dynamical map $\Phi_{t}^{12}$ can be written as a product of local dynamical
maps, i.e. $\Phi_{t}^{12}=\Phi_{t}^{1}\otimes\Phi_{t}^{2}$, if and only if  $\Lambda_{12}(t)=\kappa_{1}(t)\kappa^{\ast}_{2}(t)$ and $\kappa_{12}(t)=\kappa_{1}(t)\kappa_{2}(t)$. This means that the frequencies $\omega_{1}$ and $\omega_{2}$ are not correlated.

 We assume a Gaussian frequency distribution whose Fourier transform is obtained as
\begin{equation}\label{17}
G(\tau_{1},\tau_{2})= e^{i \omega_{0}(\tau_{1}+\tau_{2})/2-C_{11}(\tau^{2}_{1}+\tau^{2}_{2}+K \tau^{2}_{1}\tau^{2}_{2})/2},
\end{equation}
where $C_{ij}=\langle\omega_{i}\omega_{j}\rangle-\langle\omega_{i}\rangle\langle\omega_{j}\rangle$ are elements of the covariance matrix, $\langle\omega_{i}\rangle=\langle\omega_{j}\rangle=\omega_{0}/2$, and $K=C_{12}/C_{11}$ is correlation coefficient.

In order to examine Eq. (10) as a witness for initial environmental correlations, we assume two total states $\rho_{1}(0)$ and $\rho_{2}(0)$ such that $|\psi_{1}^{12}\rangle=|\psi_{2}^{12}\rangle=1/\sqrt{2}(|HH\rangle+|VV\rangle)$, and the environmental state of $\rho_{1}(0)$ is correlated whereas the environmental state of $\rho_{2}(0)$ is not. Thus, according to Eq.(10) $I\left(\rho^{S}\right)$ is always definitely positive  due to the initial environmental correlations. We have plotted  $D\left(\rho^{S}_{1}(t),\rho^{S}_{2}(t)\right)$ in terms of $\sqrt{C_{11}}\Delta n t$ for different values of the correlation coefficient in Fig. 7(a). As can be seen, the trace distance for $K=-1$, where the frequencies $\omega_{1}$ and $\omega_{2}$ are anticorrelated, gets its maximum increasing and after a specific time approaches to the value of 0.5. For $K=0$  the frequencies are not correlated  and the trace distance is always zero. The trace distance decreases after an increasing then approaches to zero for other values of K.
In Fig. 7(b), initial states of the open system are $|\psi_{1}^{12}\rangle=1/\sqrt{2}(|HH\rangle+|VV\rangle)$, and  $|\psi_{2}^{12}\rangle=\sqrt{16/18}|HH\rangle+\sqrt{2/18}|VV\rangle$ and the initial environmental states are the same as these in Fig. 7(a).
It is clear that the trace distance raises above its initial value for $K=-1$, $K=-0.99$, and $K=-0.95$, meaning that the more  anticorrelated (more distinguishable) the frequencies $\omega_{1}$ and $\omega_{2}$ are, the more information is stored outside of the open system.
It has been shown that when the frequencies $\omega_{1}$ and $\omega_{2}$ become more anticorrelated, the nature of the global dynamics becomes more non-Markovian, while  the local dynamics is Markovian\cite{Laine1}. Regarding these results with what shown in Fig. 6, it seems that time behavior of trace distance, with different initial environmental states, can be considered as a witness for determining the type of the local dynamics of the system under study. \\
\indent Finally, we conclude from Fig.7 that the trace distance may increase over its initial value due to the initial environmental correlations and the best value of the correlation coefficient is $K=-1$ for witnessing the initial correlations.\\
\section{CONCLUSIONS}
Dynamics of the trace distance with initial correlations has been studied in tripartite systems. We considered a scenario consisting of one system and two environments, and obtained a bound for the growth of distingushability in open system. The bound can be used as a witness for initial correlations among environments. The obtained inequality is general and can be applied to any interaction among three systems.
We demonstrated that initial correlations among environments under particular conditions can be witnessed by local measurements on the open quantum system. We illustrated that the bound is tight for initial classical and quantum environmental correlations. Generally, since we do not have enough  information about initial states of environments, the inequality can be applied to obtain more information about environments.\\
\indent To confirm our results we studied different tripartite systems such as a three-qubit Heisenberg XX spin chain, two Jaynes-Cummings
systems, two qubits interacting with amplitude damping environment, and an experimentally realizable example. We indicated that the distinguishability increases over its initial value due to initial correlations among the environments.\\
\indent Generalization to systems including more than three subsystems is straightforward.
\section*{ACKNOWLEDGMENTS}
\indent F.T.T. and A.S.K. would like to thank H.-P. Breuer for his 
expert advice and comments. F.T.T. also thanks J. Piilo, S.
Maniscalco, and E.-M. Laine for their useful discussions and 
comments.
%%%%%%%%%%%%%%%%%%%%%%%%%%%%%%%%%%%%%%%%%%%%%%%%%%%%%%%%%%%%%%%%%%%%%%%%%%%%%%%%%%%%%%%%%%%%%%%%%%%%%%%%%%%%%%%%%%%%%%%%%%%%%%%%%%%%%%%%%%%%%%%%%%%%%%%%%

\end{document}